\documentclass[trackchanges,twocolumn]{aastex62}
\newcommand{\V}[1]{\mathbf{#1}} 
\usepackage{amssymb}
\usepackage{amsmath}
\usepackage{hyperref}
\usepackage{natbib}
\usepackage{color}
\usepackage{graphicx}
\usepackage{amsmath}
\usepackage{bm}

\newcommand{\change}[1]{{#1}}

\newcommand{\sao}{\affiliation{Smithsonian Astrophysical Observatory, Cambridge, MA, USA}}

\newcommand{\ucb}{\affiliation{Space Sciences Laboratory, University of California, Berkeley, CA 94720-7450, USA}}

\newcommand{\uofa}{\affiliation{University of Arizona, Tucson, AZ, USA}}
\newcommand{\swri}{\affiliation{Space Science and Engineering, Southwest Research Institute, San Antonio, TX, USA}}

\begin{document}
\title{Inferred Linear Stability of Parker Solar Probe Observations using One- and Two-Component Proton Distributions}

\author[0000-0001-6038-1923]{K.G. Klein}\uofa
\author{J.L. Verniero}\ucb
\author{B. Alterman}\swri
\author[0000-0002-1989-3596]{S. Bale}
\affil{Physics Department, University of California, Berkeley, CA 94720-7300, USA}
\affil{Space Sciences Laboratory, University of California, Berkeley, CA 94720-7450, USA}
\affil{The Blackett Laboratory, Imperial College London, London, SW7 2AZ, UK}
\affil{School of Physics and Astronomy, Queen Mary University of London, London E1 4NS, UK}
\author{A. Case}\sao
\author{J.C. Kasper}\sao
\author{K. Korreck}\sao
\author{D. Larson}\ucb
\author{E. Lichko}\uofa
\author{R. Livi}\ucb
\author{M. McManus}\ucb
\author{M. Martinovi\'c}\uofa
\author{A. Rahmati}\ucb
\author{M. Stevens}\sao
\author{P. Whittlesey}\ucb

\begin{abstract}
  The hot and diffuse nature of the Sun's extended atmosphere allows
  it to persist in non-equilibrium states for long enough that
  wave-particle instabilities can arise and modify the evolution of
  the expanding solar wind. Determining which instabilities arise, and
  how significant a role they play in governing the dynamics of the
  solar wind, has been a decades-long process involving in situ
  observations at a variety of radial distances.  With new
  measurements from Parker Solar Probe (PSP), we can study what wave
  modes are driven near the Sun, and calculate what instabilities are
  predicted for different models of the underlying particle
  populations.  We model two hours-long intervals of PSP/SPAN-i
  measurements of the proton phase-space density during PSP's fourth
  perihelion with the Sun using two commonly used descriptions for the
  underlying velocity distribution. The linear stability and growth
  rates associated with the two models are calculated and compared.
  We find that both selected intervals are susceptible to resonant
  instabilities, though the growth rates and kind of modes driven
  unstable vary depending on if the protons are modeled
  using one or two components. In some cases, the predicted growth
  rates are large enough to compete with other dynamic processes, such
  as the nonlinear turbulent transfer of energy, in contrast with
  relatively slower instabilities at larger radial distances from the
  Sun.
\end{abstract}

\keywords{solar wind --- plasmas --- instabilities --- Sun: corona}


\section{Introduction}
\label{sec:intro}
Wave-particle interactions are suspected of affecting the evolution of
the solar wind as it is accelerated from the Sun's surface and expands
into the heliosphere; c.f. reviews in
\cite{Matteini:2012,Yoon:2017,Verscharen:2019}.  Such instabilities
are driven by departures from local thermodynamic equilibrium (LTE)
that are frequently modeled using velocity distributions with
anisotropic temperatures $T_{\perp,j}$ and $T_{\parallel,j}$ with
respect to local magnetic field $\V{B}$, relative field-aligned drifts between
constituent plasma populations $\Delta v_{i,j}=(\V{V}_i-\V{V}_j)\cdot
\V{B}/|\V{B}|$, and temperature disequilibrium between species $T_i
\neq T_j$.  The simultaneous effects of multiple sources of free energy
can complicate a simple linear analysis; for instance, it has been
found that the free energy contributions to unstable behavior from
different ion and electron species can be non-negligible
\citep{Chen:2016}. To address this difficulty, previous works have
applied a numerical implementation of the Nyquist instability
criterion \citep{Nyquist:1932,Klein:2017} to selected solar wind
observations from the Wind \citep{Klein:2018} and Helios
\citep{Klein:2019b} missions, finding that a majority of intervals
were unstable, including many intervals that simple parametric models
accounting for a single source of free energy would have predicted to
be stable.

Given the complexity of phase-space distributions typically found in
weakly collisional plasmas, a number of different schemes for modeling
the underlying velocity-space structure are frequently used; for
instance, it is common to treat the protons as a single, anisotropic
bi-Maxwellian or kappa distribution, or as a linear combination of
core and relatively drifting beam distributions, each with distinct
parallel and perpendicular temperatures; see the introduction of
\cite{Alterman:2018} for a review of solar wind observations of
secondary ion populations.

In this work, we select two hours-long time intervals observed by the
SPAN-i instrument from the SWEAP instrument suite
\citep{Kasper:2015} on Parker Solar Probe (PSP) \citep{Fox:2015}
during its fourth encounter with the Sun, where significant ion-scale
wave activity is observed, similar to activity previously reported in
\cite{Bowen:2020} and \cite{Verniero:2020}.  We generate both a
\textit{one-component} and \textit{two-component} model for each
measurement of the proton velocity distribution, calculating and
comparing the associated linear stability. Using the different models
produces significantly different instabilities, either in the
robustness of the associated growth rates or the kinds of waves driven
unstable. The two-component model generally predicts ion-scale waves
with characteristics more in line with the observed wave activity than
models using a single proton component. This suggests that using
overly simplistic models for ion distributions may neglect essential
kinetic-scale processes responsible for the generation of these waves,
even if these models capture macroscopic departures from LTE.

\section{Data and Methodology}
\label{sec:data}

\subsection{Parker Solar Probe Data}
\label{ssec:data}

\begin{figure}[t]
\centerline{
  \includegraphics[width=\columnwidth]{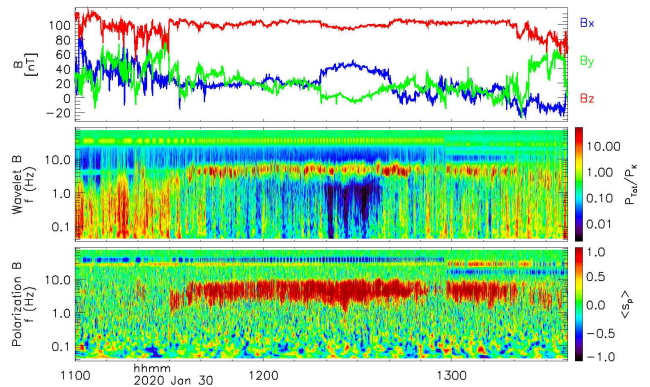}
}
\caption{Magnetic field characteristics observed by FIELDS/PSP during
  Selection A, 2020/01/30 11:00-13:30. Top row: vector components of
  $\V{B}$. Second row: Trace power spectral density normalized by
  $k^{-5/3}$ power law. Third row: Polarization of transverse magnetic
  field components, where red indicates right-handed circular
  polarization in the spacecraft frame.}
\label{fig:data_SA}
\end{figure}

\begin{figure}[t]
\centerline{
  \includegraphics[width=\columnwidth]{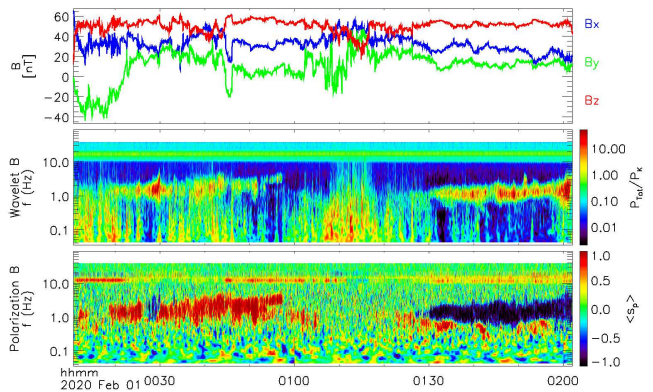}
}
\caption{Magnetic field characteristics observed by FIELDS/PSP during
  Selection B, 2020/02/01 00:10-02:00, organized in the same fashion
  as Fig.~\ref{fig:data_SA}.}
\label{fig:data_SB}
\end{figure}

We select two hours-long sections from the outbound pass of PSP's
fourth encounter with the Sun, when SPAN-i had sufficient coverage of
the proton velocity distribution to model $f_p(\V{v})$, specifically
\textbf{Selection A: 2020/01/30 11:00-13:30} (SA,
Fig.~\ref{fig:data_SA}) and \textbf{Selection B: 2020/02/01
  00:10-02:00} (SB, Fig.~\ref{fig:data_SB}). During both selections,
ion-scale electromagnetic waves are observed by the FIELDS instrument
suite \citep{Bale:2016}. Figs.~\ref{fig:data_SA} and
~\ref{fig:data_SB} show the vector magnetic field components, as well
as the trace power spectral density normalized to an \textrm{ansatz}
power-law distribution for the background turbulent spectrum of
$f^{-5/3}$, and the polarization of the transverse components of the
magnetic fields, where red (blue) indicates right-handed (left-handed)
circular polarization in the spacecraft frame. In SA, we see an
abundance of power above a $f^{-5/3}$ spectrum persist for several
hours near $3$ Hz. At the same frequencies, we see a clear signature
(red) of right-hand polarization persist for nearly the entire
duration of the more than two-hour selection.

Unlike in SA, in SB there is not a persistent signature at a nearly
constant frequency of ion-scale waves of a single handedness; both
left-handed (blue) and right-handed (red) polarized waves are
observed. There are also times during SB where no enhanced wave
activity near ion frequencies is observed.

\subsection{One- and Two-Component Proton Distributions}
\label{ssec:models}

For each $\approx 7$ second measurement where a significant fraction
of the thermal proton distribution is in the SPAN-i field of view, a
two-component fit of the observed proton energy and angle spectra is
attempted, modeling the protons as a combination of two relatively
drifting bi-Maxwellian distributions,
\begin{equation}
  f_p^{\textrm{2-comp.}}(v_\perp,v_\parallel) = \sum\limits_{j=c,b}\frac{n_j}{\pi^{3/2} w_{\perp,j}^2 w_{\parallel,j}}
  \exp\left[-\frac{v_\perp^2}{w_{\perp,j}^2}
    -\frac{\left(v_\parallel -V_{j} \right)^2}{w_{\parallel,j}^2} \right].
\end{equation}
Parallel and perpendicular are defined with respect to the local
mean-magnetic field direction, $n_j$ is the component density, $V_j$
the component bulk speed, and $w_{\perp,\parallel;j}=\sqrt{2
  T_{\perp,\parallel;j}/m_j}$ the component thermal velocities. This
fit represents our \textit{two-component model}. To mitigate the
partial FOV coverage of SPAN-i, all fitted densities were calibrated
to QTN densities. All calculations using this model are performed in
the proton center-of-mass frame.

For a model with the same macroscopic thermodynamic quantities,
i.e. total proton density as well as parallel and perpendicular
thermal pressures, that are used in a linear instability
calculation that does not represent the beam-and-core structure of the
protons observed in the inner heliosphere, we construct a
\textit{one-component model} as
\begin{equation}
  f_p^{\textrm{1-comp.}}(v_\perp,v_\parallel) = \frac{n_p}{\pi^{3/2} w_{\perp,p}^2 w_{\parallel,p}}\\
  \exp\left[-\frac{v_\perp^2}{w_{\perp,p}^2}-\frac{v_\parallel^2}{w_{\parallel,p}^2} \right].
\end{equation}
Here, the proton density is $n_p = n_c + n_b$ and the total thermal velocities are
$w_{\perp,\parallel;p}=\sqrt{2 T_{\perp,\parallel;p}/m_p}$.
We have defined the perpendicular proton temperature as
\begin{equation}
  T_{\perp,p} = \frac{n_c T_{\perp,c} + n_b T_{\perp,b}}{n_c + n_b}
\end{equation}
and the parallel proton temperature as
\begin{equation}
  T_{\parallel,p} = \frac{n_c T_{\parallel,c} + n_b T_{\parallel,b} +
    \left(\frac{n_c n_b}{n_c+n_b}\right) m_p \Delta v_{cb}^2}{n_c + n_b}.
\end{equation}
We emphasize that this is not equivalent to fitting the measured
proton VDF with a single bi-Maxwellian distribution. Our method is
employed so that both models have the same macroscopic perpendicular
and parallel proton pressures, which would not necessarily be the case
for a single bi-Maxwellian fit of protons with a significant secondary
population. 

The parameters from both models, along with measurements of the
magnetic field strength averaged to the SPAN-i measurement cadence,
are combined into the dimensionless parameters used as inputs for the
Nyquist instability analysis. We will see that the significant
differences in the underlying proton phase-space densities for the two
models lead to significant differences in the predicted unstable
behavior.

\subsection{Instability Analysis}
\label{ssec:analysis}

We employ a numerical implementation of the Nyquist instability
criterion \citep{Nyquist:1932,Klein:2017} for the hot plasma
dispersion relation for an arbitrary number of relatively drifting
bi-Maxwellian components as determined by the \texttt{PLUME} numerical
dispersion solver \citep{Klein:2015a}. The Nyquist criterion
determines the stability of a linear system of equations through a
conformal mapping of the contour integral of a dispersion relation
$\mathcal{D}(\omega,\V{k},\mathcal{P})$ over the upper-half of the
complex frequency plane. This integral counts the number of normal
mode solutions that are unstable, having $\gamma>0$, for a specific
wavevector $\V{k}$ and set of dimensionless parameters $\mathcal{P}$;
$\omega_{\textrm{r}}$ and $\gamma$ are the real and imaginary
components of the complex frequency $\omega$. Iterating this process
for multiple contours with increasing values of $\gamma$ enables the
determination of the maximum growth rate and associated
characteristics of the fastest growing mode supported by a particular
$\V{k}$. We have set $\gamma = 10^{-4} \Omega_p$ as the minimum growth
rate for a wavevector to be considered unstable. We repeat this
process over a log-spaced grid in wavevector space $k_\perp\rho_p \in
[10^{-3},3]$ and $k_\parallel\rho_p \in [10^{-2},3]$, enabling the
determination of the fastest growing mode for all wavevectors given a
particular parameter set $\mathcal{P}$.

For the one-component model, the set of dimensionless plasma
parameters is
\begin{equation}
  \mathcal{P}_{\textrm{1-comp}}=\left(\beta_{\parallel,p},\frac{w_{\parallel,p}}{c},\frac{T_{\perp,p}}{T_{\parallel,p}}
  \right)
\end{equation}
while for the two-component model, the dimensionless plasma parameters
are
\begin{eqnarray}
  \mathcal{P}_{\textrm{2-comp}}=&
  \left(\beta_{\parallel,c},\frac{w_{\parallel,c}}{c},\frac{T_{\perp,c}}{T_{\parallel,c}},
  \frac{T_{\perp,b}}{T_{\parallel,b}}, \right. \\ & \left. \frac{n_b}{n_c},
  \frac{T_{\parallel,b}}{T_{\parallel,c}}, \frac{\Delta v_{b,c}}{v_{Ac}} \right), \nonumber
\end{eqnarray}
where we define the thermal-to-magnetic pressure ratio
$\beta_{\parallel,j}=8 \pi n_j T_{\parallel,j}/B^2$, the core-proton Alfv\'en
velocity as $v_{Ac} = B/\sqrt{4 \pi m_p n_c}$,
and the speed of light $c$.  Frequencies are normalized to the proton
gyrofrequency $\Omega_p = q_p B/m_p c$.  For this study, we neglect
the contribution of alphas and other minor ions and treat the
electrons as a single isotropic distribution with density and velocity
necessary to enforce quasi-neutrality and zero net current. The impact
of the non-proton components on stability will be the focus of future
study.

\begin{figure}[t]
\centerline{
  \includegraphics[width=\columnwidth]{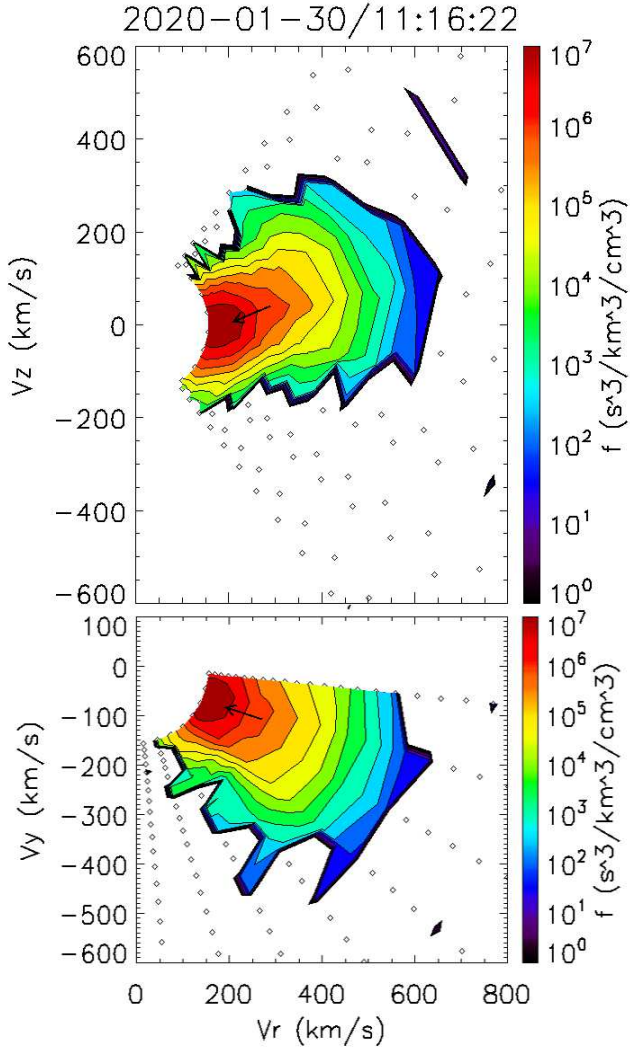}
}
\caption{SPAN-i observation of the proton velocity distribution for
  the interval under analysis in Fig.~\ref{fig:vdf_pedagogy} as a
  function of $v_z$ and $v_r$ (top) and $v_y$ and $v_r$ (bottom), in
  SPAN-i instrument co\"ordinates where $v_r =
  \sqrt{v_x^2+v_y^2}$. Diamonds represent the central values of the
  instrument's velocity space bins, color the proton distribution
  phase-space density, and {the arrow magnetic field orientation with
    the length representing the Alfv\'en speeed.} }
\label{fig:spani_example}
\end{figure}

\begin{figure*}[t]
  \includegraphics[width=\textwidth]{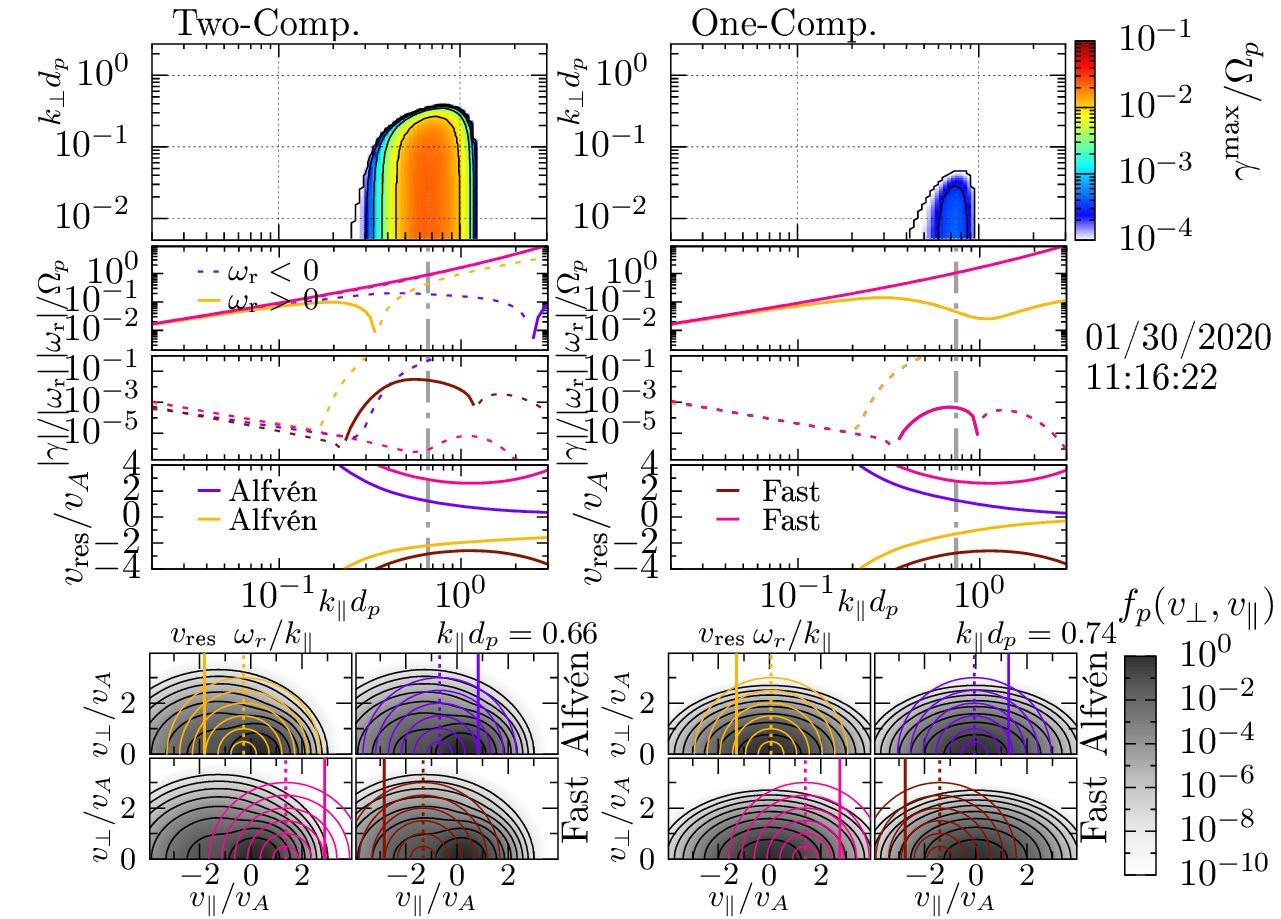}
  \caption{Comparison of linear stability and resonances for the one-
    and two-component models, left and right columns, associated with
    the SPANi observation shown in Fig.~\ref{fig:spani_example}.  Top
    row: Fastest growing mode calculated by the Nyquist method as a
    function of $k_\perp d_p$ and $k_\parallel d_p$.  Second row:
    linear dispersion relation $\omega_{\textrm{r}}(k_\parallel
    d_p)/\Omega_p$ for the four weakly damped, parallel propagating
    linear modes.  Third: Normalized growth (solid) or damping
    (dashed) rates $\gamma/\omega_{\textrm{r}}$ for the same modes.
    Fourth: Cyclotron resonant velocities normalized to $v_A$.
    Bottom: Illustration of phase (dashed vertical) and resonant
    (solid) velocities for the four modes at the wavevector associated
    with the maximum growth rate (dot-dashed line in middle panels),
    the associated curves of constant energy in the wave-frame, and
    the phase-space densities associated with the one- and
    two-component models (grey-scale).
\label{fig:vdf_pedagogy} }
\end{figure*}

Given an example SPAN-i measurement of $f_p(\V{v})$, shown in
Fig.~\ref{fig:spani_example}, both the one-component and two-component
models are constructed, producing the sets of dimensionless parameters
$\mathcal{P}_{\textrm{1-comp}}$ and $\mathcal{P}_{\textrm{2-comp}}$.
For the selected example, starting at 11:16:22 on 01/30/2020, these
sets are:
\begin{eqnarray}
  \mathcal{P}_{\textrm{1-comp}}=&
  \left(\beta_{\parallel,p}=1.0646,\frac{w_{\parallel,p}}{c}=2.786\times 10^{-4}\right.,\nonumber\\&
  \left.\frac{T_{\perp,p}}{T_{\parallel,p}}=0.389\right)
\end{eqnarray}
and
\begin{eqnarray}
  \mathcal{P}_{\textrm{2-comp}}=&
  \left(\beta_{\parallel,c}=0.410,\frac{w_{\parallel,c}}{c}=1.861\times 10^{-4},\right. \nonumber \\&
  \left. \frac{T_{\perp,c}}{T_{\parallel,c}}=0.770,
  \frac{T_{\perp,b}}{T_{\parallel,b}}=0.620, \frac{n_b}{n_c}=0.157, \right. \\ & \left.
  \frac{T_{\parallel,b}}{T_{\parallel,c}}=2.465, \frac{\Delta v_{b,c}}{v_{Ac}}=-1.350 \right). \nonumber
\end{eqnarray}
Given these sets, we calculated $\gamma^{\textrm{max}}(\V{k}
d_p)/\Omega_p$ using the Nyquist method, shown in the top two panels
in Fig.~\ref{fig:vdf_pedagogy}, which in turn allows the calculation
of $\gamma^{\textrm{max}}/\Omega_p$ over the entire wavevector range,
as well as the associated
$\omega_{\textrm{r}}^{\textrm{max}}/\Omega_p$, $k^{\textrm{max}} d_p$,
$\theta_{kB}^{\textrm{max}}$, and other eigenfunctions of the unstable
modes. For this measurement and associated models,
$\gamma^{\textrm{max}}/\Omega_p$ is significantly larger for the
two-component model and the wavevector region supporting unstable
modes is broader compared to the one-component model, though both
models predict the same mode, the parallel propagating
firehose/fast-magnetosonic wave, to be linearly unstable.

For validation, we compare these predicted properties to the normal
mode solutions for the forward and backward parallel propagating
Alfv\'en and fast-magnetosonic waves numerically calculated using the
\texttt{PLUME} dispersion solver\citep{Klein:2015a}. The central rows of
Fig.~\ref{fig:vdf_pedagogy} show the real component of the normal
mode frequency $\omega_{\textrm{r}}(k_\parallel d_p)/\Omega_p$ for
fixed $k_\perp d_p = 10^{-3}$, the normalized growth or damping rates
$\gamma(k_\parallel d_p)/|\omega_{\textrm{r}}|$, and the normalized $n
= \pm 1$ cyclotron resonant velocities,
\begin{equation}
  \frac{v_{\textrm{res}}(k_\parallel)}{v_A} =
  \frac{\omega_{\textrm{r}}(k_\parallel)-n \Omega_p}{k_\parallel v_A}
\end{equation}
where the choice of sign of $n$ is determined by the wave's
polarization and direction of propagation; $n=+1$ for the forward
Alfv\'en and backwards fast modes and $n=-1$ for the backwards
Alfv\'en and forward fast modes. For these nearly parallel modes,
there is no significant $n=0$ contribution to the wave-particle
interaction. We find good agreement with the kinds of modes and region
of wavevectors predicted to be stable and unstable from both the
Nyquist and traditional dispersion calculation.

Both models are unstable to the parallel firehose instability for this
interval, but there are significant differences---illustrated in the
bottom panels of Fig.~\ref{fig:vdf_pedagogy}--- in the resonant
coupling between the protons and the electric field.  The wave-phase
velocity for each of the four parallel propagating modes at a fixed
wavevector $k_\parallel d_p$, set to be $|k^{\textrm{max}}|d_p$ for
the one- or two-component model, is illustrated as a dashed vertical
line compared to the model phase-space density
$f_p(v_\perp/v_A,v_\parallel/v_A)$. The $n=\pm 1$ cyclotron resonant
velocity is shown as a solid vertical line, and contours of constant
energy in the wave-frame are illustrated as colored half-circles. The
sign of the pitch angle gradient of $f_p$ where the resonant velocity
meets the contours of constant energy determines if energy is
transferred from the wave to the protons, leading to damping of the
wave, or from the protons to the wave, leading to excitation and
instability. For this interval, the fitting of a secondary proton
population leads to the suppression of the unstable anti-beam-aligned
fast mode and the enhancement of the beam-aligned fast mode's growth
rate. The beam component also significantly increases the damping rate
of the anti-beam aligned Alfv\'en mode, leading it to switch
propagation directions at $k_\parallel d_p \approx 0.3$.

\section{Inferred Stability Across Selections}
\label{sec:results}

\begin{figure}
  \includegraphics[width=0.5\textwidth]{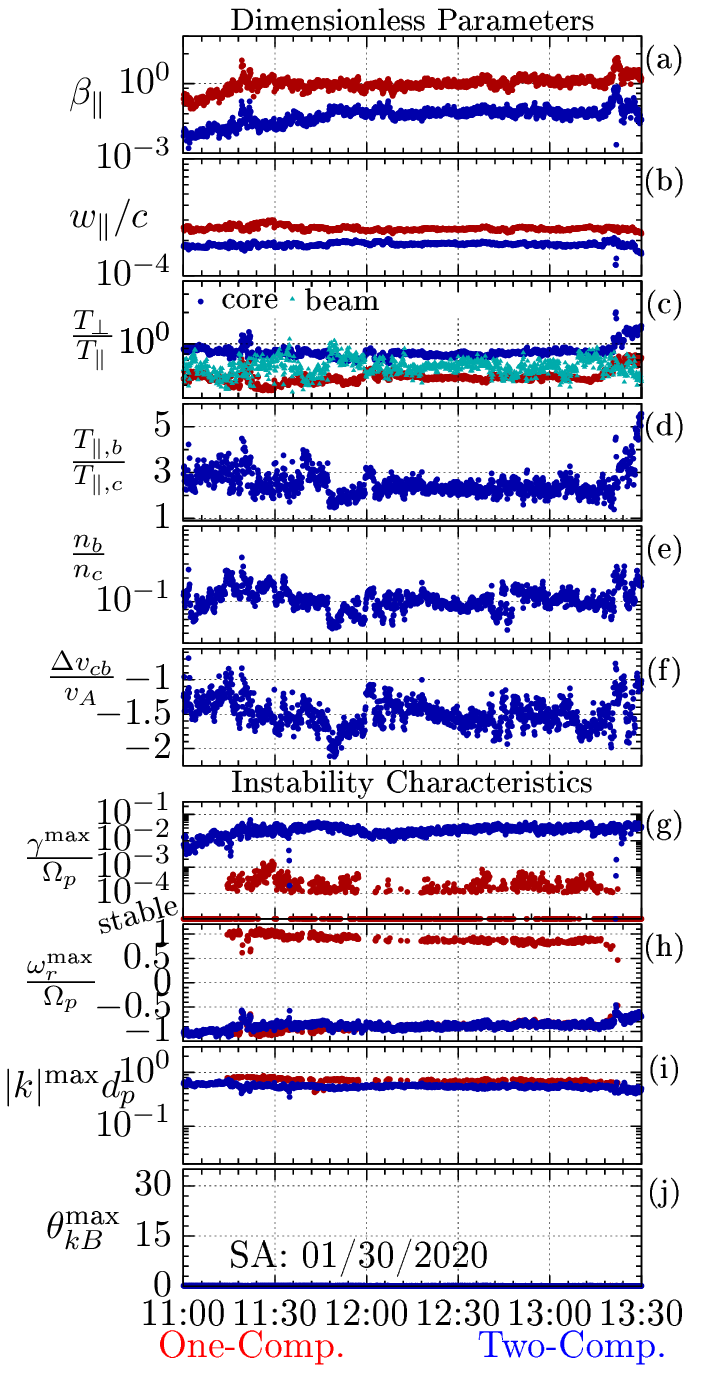}
\caption{Dimensionless parameters from one- and two-component (red and
  blue) models for SA, (a-f) and calculated instability
  characteristics, (g-j).  (a): thermal-to-magnetic pressure ratio
  $\beta_{\parallel,p}$ or $\beta_{\parallel,c}$, (b): thermal-speed
  ratio $w_{\parallel,p}/c$ or $w_{\parallel,c}/c$, (c): temperature
  anisotropy $T_{\perp,p}/T_{\parallel,p}$ or
  $T_{\perp,c}/T_{\parallel,c}$ ($T_{\perp,b}/T_{\parallel,b}$ in
  teal), (d): temperature disequilibrium
  $T_{\parallel,b}/T_{\parallel,c}$, (e): density ratio $n_b/n_c$,
  (f): relative drift velocity $\Delta v_{bc}/v_A$. (g): maximum
  growth rate $\gamma^{\textrm{max}}/\Omega_p$ (h): normal mode real
  frequency $\omega_{\textrm{r}}^{\textrm{max}}/\Omega_p$ (i \& j):
  Amplitude and angle, $|k|^{\textrm{max}} d_p$ and
  $\theta_{kB}^{\textrm{max}}$ of the wavevector associated with
  fastest growing mode.
\label{fig:parameters.SA} }
\end{figure}

\begin{figure}[t]
  \includegraphics[width=0.5\textwidth]{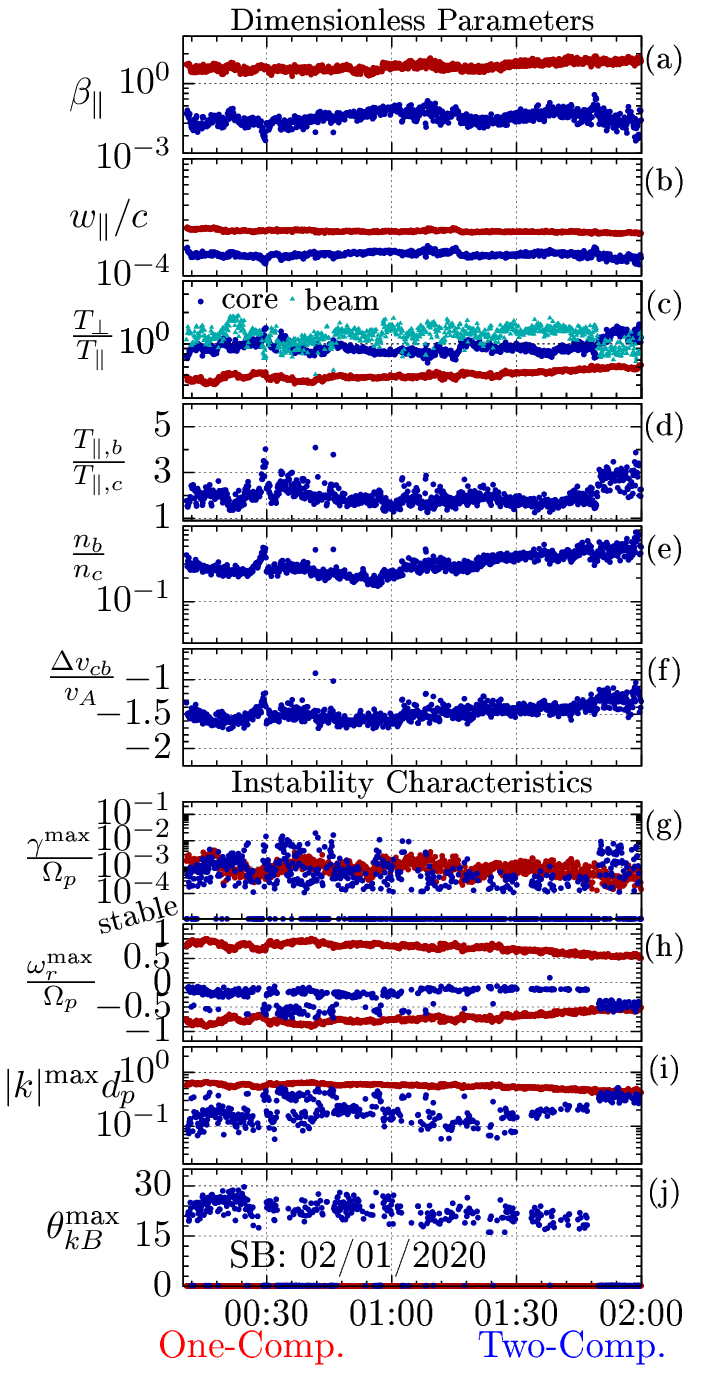}
\caption{Dimensionless parameters and calculated instability
  characteristics from one- and two-component (red and blue) models for
  SB organized in the same format as Fig.~\ref{fig:parameters.SA}.
\label{fig:parameters.SB} }
\end{figure}

The Nyquist instability analysis described in \S \ref{ssec:analysis} is
performed over the entirety of SA, Fig.~\ref{fig:parameters.SA}, and
SB, Fig.~\ref{fig:parameters.SB}, for both the one- and two-component
models (red and blue).

For both selections, we see different predicted unstable behavior for
the two models. Using the one-component model for SA, only $40.5\%$ of
the intervals are found to be unstable, and of those most have
relatively weak growth rates, with a median value of
$\bar{\gamma}^{\textrm{max}}_{1-\textrm{comp}} = 2.33^{3.41}_{1.57}
\times 10^{-4} \Omega_p$. The sub- and super-scripts represent the
$25^{\textrm{th}}$ and $75^{\textrm{th}}$ percentiles of the unstable
mode growth rate distribution. These are parallel firehose
instabilities, where sufficiently extreme parallel-to-perpendicular
thermal pressure ratios, manifest in a one-component proton
distribution, change the sign of the velocity gradient at the
cyclotron resonant velocity such that energy is extracted from the
protons to drive an unstable fast-magnetosonic mode. Due to the
symmetry of the one-component model, both forward and backward
propagating modes are driven. No other kinds of unstable modes are
supported by the one-component model during SA.

For the two-component model, $99.9\%$ of the intervals in SA are found
to be unstable, with a median growth rate of
{$\bar{\gamma}^{\textrm{max}}_{2-\textrm{comp}}= 2.54^{3.24}_{1.92}
  \times 10^{-2} \Omega_p$}, two orders of magnitude larger than for
the one component model. All of the unstable intervals are associated
with parallel propagating fast-magnetosonic modes with
$|k|^{\textrm{max}} d_p \approx 0.5$. Unlike the symmetrically emitted
unstable waves from the one-component model, the unstable modes from
the two-component model only propagate in the same direction as the
secondary proton population.\footnote{We define the radial component of our
  co\"ordinate system to align with the mean magnetic field. In both
  SA and SB, PSP was in a region of Sunward magnetic polarity, meaning
  that the anti-Sunward propagating secondary proton populations have
  a negative velocity with respect to the primary proton
  population.}  The maximum growth rate of the unstable fast mode is
enhanced due to an increased phase-space density associated with the
secondary proton population, while the anti-beam aligned fast-mode
resonance is effectively starved of protons with which to interact,
leading to damping rather than instability for this mode.

We find differences in the kinds of instabilities predicted for the
two models in SB. Ninety-nine percent of the intervals are predicted
to be linearly unstable to the parallel propagating firehose
instability for the one-component model, with a median growth rate of
{$\bar{\gamma}^{\textrm{max}}_{1-\textrm{comp}}= 9.26^{13.3}_{6.12}
  \times 10^{-4} \Omega_p$}.  This is \textit{not} the case for the
two-component model. The median growth rate for the two-component
model is similar, {$\bar{\gamma}^{\textrm{max}}_{2-\textrm{comp}}=
  6.43^{15.6}_{3.30} \times 10^{-4} \Omega_p$}, however only
$55.7\%$ of the intervals are found to be unstable and the associated
fastest growing mode oscillates between a beam-aligned, parallel
propagating firehose mode and an oblique instability. This
demonstrates that fitting a secondary component does not universally
enhance the predicted growth rate and that more sophisticated
treatments of velocity-space structure can lead to the generation of
different kinds of unstable modes.

\begin{figure}[t]
  \includegraphics[width=\columnwidth]{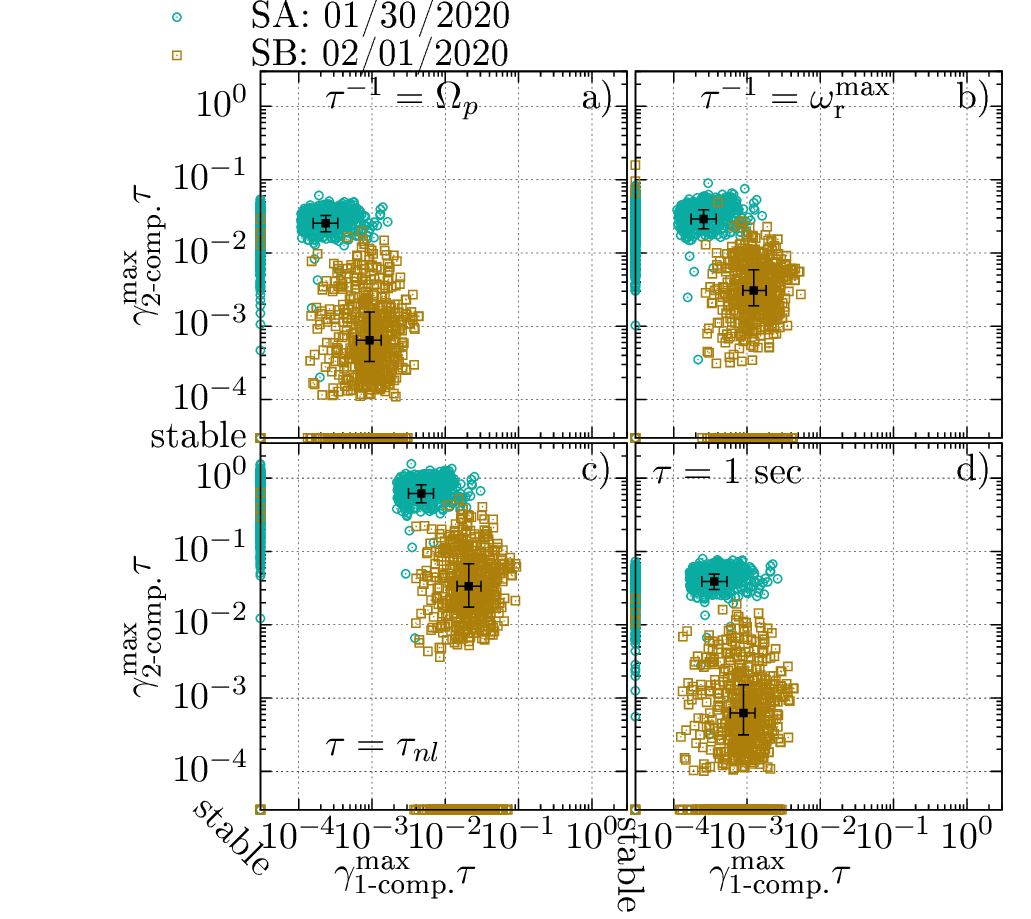}
\caption{Comparison of normalized growth rates for the two models for
  SA (teal) and SB (gold), with the abscissa and ordinate mapping the
  one- and two-component rates. In panels a,b,c, and d,
  $\gamma^{\textrm{max}}$ is normalized to $\Omega_p$,
  $\omega_{\textrm{r}}^{\textrm{max}}$, $\tau_{nl}^{-1}$ and $1$ Hz
  respectively. Black dots and bars correspond to medians and
  $25^{\textrm{th}}$ and $75^{\textrm{th}}$ percentiles associated with the unstable intervals.
\label{fig:gamma_compare} }
\end{figure}

As seen in Fig.~\ref{fig:gamma_compare}, $\gamma^{\textrm{max}}/\Omega_p$
is generally larger for the two-component model than for the
one-component model for SA. This is not the case for SB, where
more of the one-component intervals are unstable, while the variance
in the growth rate for the two-component model is larger. When
re-normalized to the normal mode frequency
$\omega_{\textrm{r}}^{\textrm{max}}$, Fig~\ref{fig:gamma_compare}b, we
see an enhancement in the growth rates for the two-component model in
SB, while the the other growth rates remain relatively unaffected.

Other time scales of potential interest include an estimate for the
non-linear cascade rate at the wavevector of fastest growth,
\begin{eqnarray}
  \gamma^{\textrm{max}}\tau_{nl}= &
    \left(\frac{\gamma^{\textrm{max}}}{v_A}\right)
  \left(k_{\textrm{break}}\right)^{-1/3}
  (|\V{k}^{\textrm{max}}|)^{-2/3}\\
=&\left(\frac{\gamma^{\textrm{max}}}{\Omega_p}\right)
  (\frac{2 \pi f_{\textrm{break}}}{\Omega_p} \frac{v_A}{v_{sw}})^{-1/3}
  (|\V{k}^{\textrm{max}}d_p|)^{-2/3} \nonumber
\end{eqnarray}
where we approximate the transition from the injection to the inertial
ranges of turbulence as $k_{\textrm{break}} = 2 \pi
f_{\textrm{break}}/v_{sw}$ with $f_{\textrm{break}}$ found to be
approximately $10^{-3}$ Hz when constructing trace power-spectral
density curves for either SA or SB, not shown. These values are in
rough agreement with the results reported in \cite{Chen:2020}.  The
cascade time is estimated as the critically balanced nonlinear cascade
rate, $\tau_{nl} \sim
\omega_{\textrm{Alfv\'en}}^{-1}$\citep{Goldreich:1995,Mallet:2015}.
Previous analysis between 0.3 and 0.7 au \citep{Klein:2019b} found
that $\gamma^{\textrm{max}}$ never exceeded the estimated nonlinear
cascade rate, though the two rates were found to be within an order of
magnitude, with $50\%$ of the intervals having
$\gamma^{\textrm{max}}\tau_{nl} \gtrsim 0.2$. For the two-component
model in SA, the maximum growth rate is of the same order as
$\tau_{nl}^{-1}$, with a median value of
$\bar{\gamma}^{\textrm{max}}_{2-\textrm{comp}}\tau_{nl}=0.618^{0.813}_{0.463}$,
indicating that these predicted instabilities operate on similar
timescales as the nonlinear transport of energy through these spatial
scales. Importantly, while
$\bar{\gamma}^{\textrm{max}}_{2-\textrm{comp}} \sim \tau_{nl}^{-1}$,
the median value of
$\bar{\gamma}^{\textrm{max}}_{1-\textrm{comp}}\tau_{nl}$ is $4.70^{6.90}_{3.14}\times
10^{-3}$ for the same interval. This emphasizes that our choice of
different models for the proton phase-space density will lead to
drastically different interpretations of the importance of different
physical processes. The impact of these instabilities, especially when
the ions are modeled as multiple components, on the turbulent
transport of energy must be considered in future modeling efforts.
The median values of $\gamma^{\textrm{max}}\tau_{nl}$ are comparable
for SB, with $\bar{\gamma}^{\textrm{max}}_{1-\textrm{comp}}\tau_{nl} =
2.08^{3.04}_{1.45} \times 10^{-2}$ and
$\bar{\gamma}^{\textrm{max}}_{2-\textrm{comp}}\tau_{nl} = 3.55^{6.81}_{1.74} \times
10^{-2}$, again showing that the two-component model does not
universally enhance growth rates compared to the one-component
model. To remove variations associated with the normalization by
$\Omega_p$ due to changes in $|\V{B}|$ as a function of time, we also
plot the growth rate in Hertz, Fig~\ref{fig:gamma_compare}d, and see a
distribution of growth rates similar to that seen in panel a.

By design, the one- and two-component models have the same
parallel and perpendicular thermal pressures for a given interval,
which can be characterized by the firehose \citep{Kunz:2015}
\begin{equation}
\Lambda_F=\frac{\beta_\parallel - \beta_\perp}{2} + \frac{\sum_j n_j m_j |\Delta \tilde{v}_j|^2}{\sum_j(n_j m_j) v_A^2}
\end{equation}
or mirror \citep{Hellinger:2007}
\begin{equation}
\Lambda_M=\sum_j \beta_{\perp,j} \left(\frac{\beta_{\perp,j}}{\beta_{\parallel,j}}-1 \right) - \frac{\left(\sum_j q_j n_j\frac{\beta_{\perp,j}}{\beta_{\parallel,j}}\right)^2}{2 \sum_j\frac{(q_j n_j)^2}{\beta_{\parallel,j}}}
\end{equation}
criterion, where $\Delta \tilde{v}_j$ is the difference between the bulk speed
of component $j$ and the center of mass velocity.  When these
criterion exceed unity, large-scale firehose or mirror instabilities
are generated.  For both SA and SB, the amplitude of neither criteria
exceeds $\sim 0.5$ for either model; therefore, it is the resonances
between the proton distribution and the associated electromagnetic
fields and not the excess macroscopic parallel or perpendicular
pressures that drives the predicted unstable wave modes.

\begin{figure*}[t]
  \includegraphics[width=\textwidth]{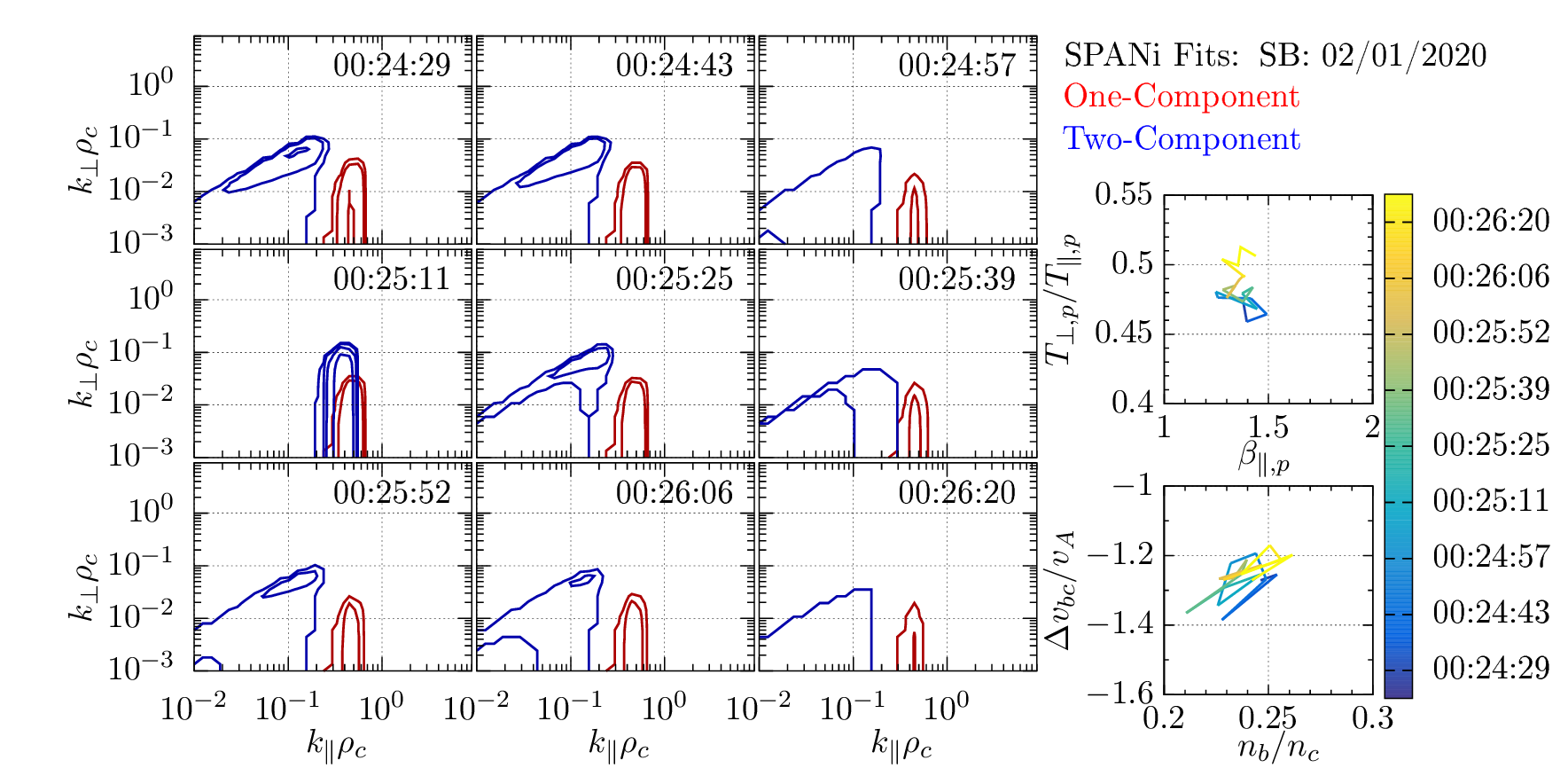}
\caption{Left: Contours of constant $\gamma^{\textrm{max}}/\Omega_p$
  as a function of $k_\perp \rho_c$ and $k_\parallel \rho_c$ for the
  one- and two-component models (red and blue) for nine intervals at
  the start of SB. Right: Temporal variation of
  $T_{\perp,p}/T_{\parallel,p}$ and $\beta_{\parallel,p}$ from the
  one-component model (top) and of the relative
  drifts and densities of the two-component model (bottom).
\label{fig:time_variation} }
\end{figure*}

Slight changes in the relative drift speed between the two proton
populations and their densities can have a significant impact on the
kind of unstable mode predicted to be generated.  This is illustrated
in Fig.~\ref{fig:time_variation}, where nine sequential illustrations
of contours of constant $\gamma^{\textrm{max}}(\V{k})$ are shown for
the one- and two-component models for SPAN-i observations from near the
beginning of SB. Throughout these two minutes both the maximum growth
rates and regions of unstable wavevectors are largely unchanged for
the one-component model.  This is expected given that
$T_{\perp,p}/T_{\parallel,p}$ and $\beta_{\parallel,p}$ are relatively
constant over this time, remaining consistent with a parallel
propagating firehose instability.  For the two-component model,
oblique modes are initially driven. A minute into the
sequence, the maximum growth rate transitions to a parallel
propagating wavevector, and then transitions back to an oblique
instability. These transitions correspond to a temporary dip in the
relative density of the beam component and an increase in the relative
drift speed. Given that many kinds of waves are observed in this
section of data, it appears plausible that these transitions between
parallel and oblique instabilities may be real, but are not properly
accounted for in overly simplistic models of the protons as a single
anisotropic distribution, which only drive one kind of unstable mode.

\begin{figure}[t]
  \includegraphics[width=\columnwidth]{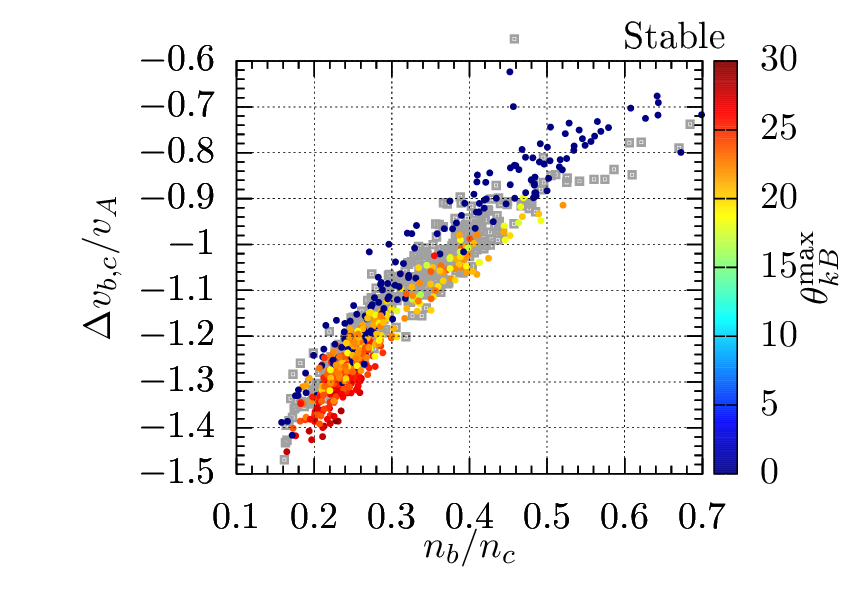}
\caption{Wavevector angle $\theta_{kB}^{\textrm{max}}$ of fastest growing mode during
  SB calculated using the two-component model, indicated by color, as
  a function of relative densities $n_b/n_c$ and drift velocities
  $\Delta v_{b,c}/v_A$. Grey squares indicate intervals predicted to
  be linearly stable.
\label{fig:nb_dv} }
\end{figure}

We note that there is not a simple parametric function dependence only
on $n_b/n_c$ and $\Delta v_{b,c}/v_A$ that divides the parallel
unstable modes from the oblique modes. In Fig.~\ref{fig:nb_dv}, we
plot the angle of the fastest growing mode
$\theta_{kB}^{\textrm{max}}$ for the two-component model for SB as a
function of these two parameters.  Generally, the larger the relative
drift, the more likely the model is predicted to generate an oblique
unstable mode, with the transition between parallel and oblique modes
arising at lower drifts for larger relative beam densities. However,
we find many stable intervals with very similar drifts and densities
to the intervals unstable to the generation of both parallel and
oblique unstable modes. This can be understood by recalling that the
variation of the temperatures and anisotropies of the individual
proton components will have a significant impact on the predicted
stability of the system that is not captured in this reduced parameter
space. Due to this complexity, we do not attempt to offer a simple
parametric prescription for this transition between parallel and
oblique instabilities in this work, but do note again that if this
distribution is treated as a single proton population, the only
instability supported is the parallel-propagating, fast/magnetosonic
firehose instability.

\section{Conclusions}
\label{sec:discussion}

In this work, we have selected two hours-long intervals where in situ
measurements of the local plasma conditions have been made during
PSP's fourth perihelion orbit. These measurements coincide with
significant ion-scale wave activity as observed by the FIELDS
magnetometers. The proton phase-space densities have been modeled as
either a single anisotropic population, or two relatively drifting
anisotropic populations. The linear stability of both models was
calculated, with strikingly different predictions for the supported
linear modes. In the first selection, both models produce the same kind of
unstable mode, but the two-component model drives instabilities that
grow nearly two orders of magnitude faster, fast enough to potentially
act on the same timescales as the local nonlinear turbulent transfer of
energy. Additionally, the two-component model for SA only drives
instabilities propagating in a single direction, as opposed to the
one-component model where waves are driven both Sunward and
anti-Sunward due to the enforced symmetry of the simplified
description of the protons. For the second selection, modeling the
protons using two components does not make the plasma more unstable,
but does change the kind of unstable modes driven, leading to an
oscillation between the production of parallel and oblique propagating
waves.

\change{As future lines of inquiry, we intend on extending this work to
investigate the predicted growth rates and waves concurrently observed
with other plasma parameters and solar wind conditions, such as
intervals where the total parallel proton pressure is exceeded by the
total perpendicular pressure. We will also include additional sources of
free energy associated with minor ions and electrons, to determine if
they act to enhance or stabilize these growing modes. This work will
help to ascertain under what conditions which models may suffice to
properly describe kinetic processes.} Importantly, as the instabilities under consideration are
resonant, we must also consider the impact of departures from
bi-Maxwellian distributions, either using other analytic
prescriptions, e.g. kappa \citep{Livadiotis:2015} or flattop
distributions\citep{Klein:2016,Wilson:2020c}, or via a direct
numerical integration of the observed phase-space density
\citep{Verscharen:2018}.


The SWEAP Investigation and this publication are supported by the PSP
mission under NASA contract NNN06AA01C. K.G.K. is supported by NASA
ECIP Grant 80NSSC19K0912. An allocation of computer time from the UA
Research Computing High Performance Computing at the University
of Arizona is gratefully acknowledged.

\bibliographystyle{apj}


\end{document}